\documentclass[11pt,showpacs]{revtex4-1}

\usepackage{epsfig}
\input epsf.tex
\usepackage{amssymb}
\usepackage{amsmath}
\usepackage{amsthm}
\usepackage{amsopn}

\def\comment#1{}

\def\beq{\begin{equation}}
\def\eeq{\end{equation}}
\def\bea{\begin{eqnarray}}
\def\eea{\end{eqnarray}}

\usepackage{ulem}
\usepackage{cancel}
\usepackage{color}

\begin{document}

\title{Circular polarization from linearly polarized laser beam collisions}

\author{Rohoollah Mohammadi$^{1}$\footnote{rmohammadi@ipm.ir}, I. Motie$^{2}$  and She-Sheng Xue$^{3}$\footnote{xue@icra.it}}
\affiliation{$^{1}$ School of physics, Institute for research in fundamental sciences (IPM), Tehran, Iran.\\
$^2$
Department of Physics, Mashhad Branch, Islamic Azad University, Mashhad, Iran.\\
$^{3}$
ICRANet, P.zza della Repubblica 10, I--65122 Pescara, Physics Department, University of Rome {\it La Sapienza}, P.le Aldo Moro 5, I--00185 Rome, Italy.}
%\affiliation{$^{(b)}$Department of Physics, Isfahan University of Technology, Isfahan 84156-83111, Iran}
%\maketitle

\begin{abstract}
To probe the nonlinear effects of photon-photon interaction in the quantum electrodynamics, we study the generation of circular polarized photons by the collision of two linearly polarized laser beams. In the framework of the Euler-Heisenberg effective Lagrangian and the Quantum Boltzmann equation for the
time evolution of the density matrix of polarizations, we calculate the intensity of circular polarizations generated by the collision of two linearly polarized laser beams and estimate the rate of generation. As a result, we show that the generated circular polarization can be experimentally measured, on the basis of optical laser beams of average power KW,
%beam diameter $\sim {\rm cm}$ and wavelength $\sim 1{\rm \mu m}$,
which are currently available in laboratories. Our study presents a valuable supplementary  to other theoretical and experimental frameworks to study and measure the nonlinear effects of photon-photon interaction in the quantum electrodynamics.
\end{abstract}

%\pacs{13.15.+g,34.50.Rk,13.88.+e}

\maketitle
%\newpage

\section{Introduction}
Due to the fact that in classical electrodynamics Maxwell's equations are linear, the light by light scattering in the vacuum does not occur, instead they obey superposition.  In the context of the quantum electrodynamics (QED), the specific process of photon-photon scattering is present, as the result of a virtual electron-positron pair production by the two initial photons, followed by the annihilation of this pair into the final photons, for more detailed description, see for example Ref.~\cite{lif}. These nonlinear interactions are described by the effective Lagrangian of Euler and Heisenberg \cite{euler} (see review articles \cite{revs}). This effective Lagrangian modifies Maxwell's equations for the average values of the electromagnetic quantum fields \cite{rev1} and affects the properties of the QED vacuum \cite{rev2}. In addition,
based on the Euler and Heisenberg effective Lagrangian, the effects of the photon splitting, QED birefringence and dichroism were studied (see Refs.~\cite{adler1971,hh1997}).

For many years, the
search of these non-linear QED effects has been restricted to projected
particle physics experiments with accelerators. The difficulties of the measuring these effects stem from the smallness of nonlinear interaction of order ${\mathcal O}(\alpha^4)$, where the fine-structure constant $\alpha=1/137$.
Nevertheless, these non-linear QED effects in the vacuum will possibly
become testable at energy densities achievable with
ultra-high power lasers in the near future. Based on recent
advanced laser technologies, there are many ongoing
experiments: x-ray free-electron laser (XFEL) facilities
\cite{XFEL}, optical high-intensity laser facilities such as Vulcan
\cite{vulcan}, petawatt laser beam \cite{peta} and ELI \cite{eli}, as well as SLAC E144 using nonlinear Compton scattering \cite{burke1997}; for details, see Refs.~\cite{Ringwald,mtb2006,laser_report}. This leads to the
physics of the ultrahigh-intensity laser-matter interactions in
the critical field \cite{keitel2008}.

Recently,  the nonlinear effect of the  photon-photon interactions is
shown to manifest itself in a variety of the ways such as (i) a phase shift in intense laser beams crossing one another \cite{rev4}, (ii) a frequency shift of a photon propagating in an intense laser \cite{rev5}, (iii) the polarization
effects of QED vacuum birefringence and dichroism in crossing laser beams \cite{rev6,rev7,rev8} and (iv) the photon-photon scattering in collisions of laser pulses \cite{king}, where it is shown that a single $10$ PW
optical laser beam splitting into two counter-propagating pulses is sufficient for measuring the elastic
process of photon-photon scattering, moreover, when these pulses are sub-cycle, results suggest that the inelastic process of photon-photon scattering should
be measurable too. It should be mentioned that PW
optical laser beams are required to measure these effects of photon-photon scattering, the reason will be given in the next section.

In this Letter, however, we attempt to study the effect of QED birefringence
in the collision of laser beams.
We show that in the collision of two linearly polarized laser beams, the circular polarization can be generated by nonlinear QED effects of the photon-photon interaction, i.e., the Euler-Heisenberg effective Lagrangian in the vacuum. It is important to point out that the rate of generating circular polarization is large enough to be experimentally measured for the collision of two KW optical laser beams, which have already been  achieved in laboratories nowadays. The reason will be given in the concluding section.
%have the beam intensities $\sim {\rm KW}/cm^2$, beam diameter $\sim {\rm cm}^2$ and wavelength $\sim 1{\rm \mu m}$
We recall that the generation of circular polarizations for Cosmic Background Microwave (CBM) radiation due to the Euler-Heisenberg effective Lagrangian was discussed in Ref.~\cite{xuei}.

%%%%%%%%%%%%%%%%%%%%%%%%%%%%%%%%%%%%%%%%%%%%%%%%%%%%%%%%%%%%%%%%%%%%%%%%%%%%%%%%%%%%%%%%%%%%%%%%%%%%%%%%%%%%%%%%%%%%%%%%%%%%%%%%%%%%%%%%
\section{The rate of photon-photon scattering}\label{c-rate}
%%%%%%%%%%%%%%%%%%%%%%%%%%%%%%%%%%
The photon-photon (light by light) scattering (in the vacuum) is a special process of quantum electrodynamics (QED), which does not occur in classical electrodynamics, owing to the fact that Maxwell's equations are linear. The leading contribution to the photon-photon scattering comes from Feynman ``box'' diagrams of the four external photon lines, which is the leading term in the Euler-Heisenberg effective Lagrangian. The total cross-section of photon-photon scattering $\sigma_{\gamma\gamma}$ is given by (see Ref.~\cite{lif})
%Fig.~\ref{cross3} (see Ref.~\cite{lif}) and
\begin{eqnarray}\label{cross1}
    \sigma_{\gamma\gamma}&\simeq &0.031\alpha^4\lambda_e^2\left(\frac{\omega}{m_e}\right)^4,\quad \omega\ll m_e,\label{cross1}\\
		  \sigma_{\gamma\gamma}&\simeq & 4.7\alpha^4\lambda_e^2\left(\frac{m_e}{\omega}\right)^2,\quad \omega\gg m_e\label{cross2}
\end{eqnarray}
where $\omega$, $m_e$ and $\lambda_e$ are the photon energy in the center-of-mass system,
electron mass and Compton wavelength, the maximal cross-section  is around $\omega\approx m_e$.
Using the cross-section of photon-photon scattering and intensities of laser beams available in laboratories, we estimate the rate of laser light-light scattering as follow
\begin{equation}\label{rate0}
    R_{\gamma\gamma}=\sigma_{\gamma\gamma}\, n^{in}_{\gamma}\,N^T_\gamma,
\end{equation}
where $N^T_\gamma$ is the number of (target) laser-photons interacting with (incident) laser-photons and $n^{in}_{\gamma}$ is the number of
(incident) laser-photons per second and per cross-sectional interacting area $A$ ($\sqrt{A}\sim$ being the size of laser-beams). These quantities can be written as
 \begin{equation}\label{rate1}
    n^{in}_{\gamma}= \frac{Q^{T}(\omega)}{\omega},\,\,\,\,\,\,\,\,\,\,\,N^T_\gamma=\frac{Q^{T}(\omega')}{\omega'}A\,\Delta t,
 \end{equation}
where $Q^{T}(\omega')$ is the intensity of (target) laser-beam and $Q^{in}(\omega)$ is the intensity of (incident) laser-beam, $\Delta t$ represents the time-interval of two laser beams interacting.
Eqs.~(\ref{rate0}) and (\ref{rate1}) lead to
\begin{equation}\label{rate2}
    R_{\gamma\gamma}=\sigma_{\gamma\gamma}\, A\,\Delta t\, \left(\frac{Q^{T}(\omega')}{\omega'}\right)\left(\frac{Q^{in}(\omega)}{\omega}\right).
\end{equation}
In the case of $\omega=\omega'\sim {\rm eV}$, the total cross-section of Eq.~(\ref{cross1}) is very small, $\sigma_{\gamma\gamma}\sim 10^{-54}cm^2$. Assuming that the intensities $Q$ of incident and target laser beams are equal, then we obtain
\begin{equation}\label{rate3}
   R_{\gamma\gamma} \simeq \frac{\sigma_{\gamma\gamma}}{A}\left(\frac{\bar{P}}{\omega}\right)^2\,\left(\frac{\Delta t}{ \rm sec}\right)\,\left(\frac{\omega}{\rm eV}\right),
\end{equation}
where the average power of laser beams $\bar{P}=A\,Q^{in}=A\,Q^{T}$. Due to the smallness of cross-section $\sigma_{\gamma\gamma}$ in Eq.~(\ref{rate3}), in order to observe a scattered photon per second ($R_{\gamma\gamma}\sim 1$), one needs
petawatt laser beams such as Vulcan laser $\sim10$ PW  \cite{vulcan} and Petawatt laser \cite{peta}, where the laser duration time $\tau\sim {\rm fs}$, laser beam diameter $\sqrt{A}\sim 50\,{\rm cm}$, repetition rate $f_{{\rm pulse}}\sim 1/{\rm sec}$ and energy per pulse $\varepsilon_{{\rm pulse}}\sim {\rm kJ}$. This agrees with the result reported in Ref.~\cite{king}. This explains the reason why PW laser beams are required to measure the effects of nonlinear photon-photon scattering.

%%%%%%%%%%%%%%%%%%%%%%%%%%%%%%%%%%%%%%%%%%%%%%%%%%%%%%%%%%%%%%%%%%%%%%%%%%%%%%%%%%%%%%%%%%%%%%%%%%%%%%%%%%%%%%%%%%%%%%%%%%%%%%%%%%%%%%
\section{ Euler-Heisenberg Lagrangian and circular polarizations}
%%%%%%%%%%%%%%%%%%%%%%%%%%%%%%%%%%%%%%%%%%%%%%%%%%%%%%%%%%%%%%%%%
We attempt to study the generation of the circularly polarized photons due to the Euler-Heisenberg effective
Lagrangian in the collision of two linearly polarized laser beams. The Euler-Hesinberg
effective Lagrangian is given by \cite{euler}:
\begin{eqnarray}
\pounds_{eff}=
-\frac{1}{4}F_{\mu\nu}F^{\mu\nu}+\frac{\alpha^2}{90m_e^4}
\left[(F_{\mu\nu}F^{\mu\nu})^2
+\frac{7}{4}(F_{\mu\nu}\tilde{F}^{\mu\nu} )^2\right],
\label{eh}
\end{eqnarray}
where the first term $\frac{1}{4}F_{\mu\nu}F^{\mu\nu}$ is the classical Maxwell
Lagrangian. Although the Euler-Heisenberg effective Lagrangian was obtained for constant electromagnetic fields, we approximately use it to represent the interaction of laser fields in the following calculations.
We express the electromagnetic field strength
$F_{\mu\nu}=\partial_\mu A_\nu-\partial_\nu A_\mu$, the dual field strength $\tilde F^{\mu\nu}
=\epsilon^{\mu\nu\rho\sigma}F_{\rho\sigma}$ and the gauge field $A_\mu$ in terms of plane wave
solutions in the Coulomb gauge \cite{zuber},
\begin{eqnarray}
A_\mu(x) = \int \frac{d^3 k}{(2\pi)^3
2 k^0} \left[ a_i({\bf k}) \epsilon _{i\mu}({\bf k})e^{-ik\cdot x}+
a_i^\dagger ({\bf k}) \epsilon^* _{i\mu}({\bf k})e^{ik\cdot x}
\right],
\end{eqnarray}
where $\epsilon _{i\mu}({\bf k})=(0,\vec{\epsilon} _{i}({\bf k}))$ are the
polarization four-vectors and the index $i=1,2$, representing two
transverse polarizations of a free photon with four-momentum $k$
and $k^0=|{\bf{k}}|$. $a_i({\bf k})$ and $a_i^\dagger ({\bf k})$ are creation
and annihilation operators, which satisfy the canonical
commutation relation
\begin{equation}
\left[  a_i ({\bf k}), a_j^\dagger ({\bf k}')\right] = (2\pi )^3
2k^0\delta_{ij}\delta^{(3)}({\bf k} - {\bf k}' ).
\label{comm}
\end{equation}
The number operator $ D^0_{ij}({\bf k})\equiv
a_i^\dag ({\bf k})a_j({\bf k})$.

An ensemble of photons in a general mixed state
is described by a normalized density matrix $\rho_{ij}\equiv
(\,|\epsilon_i\rangle\langle \epsilon_j|/{\rm tr}\rho)$, $\rho_{ij}({\bf p})$ is the general density-matrix
in the space of polarization states with a fixed
energy-momentum ``${\bf p}$'' per unit volume,
the dimensionless expectation values for Stokes parameters are given
by
\begin{eqnarray}
I\equiv\langle  \hat I \rangle &=& {\rm tr}\rho\hat I%=\rho_{11}+\rho_{22}
=1,\label{i}\\
Q\equiv\langle  \hat Q \rangle &=& {\rm tr}\rho\hat{Q}=\rho_{11}-\rho_{22},\label{q}\\
U\equiv\langle
 \hat U\rangle &=&{\rm tr}\rho\hat{U}=\rho_{12}+\rho_{21},\label{u}\\
V\equiv\langle  \hat V \rangle &=& {\rm
tr}\rho\hat{V}=i\rho_{21}-i\rho_{21},
\label{v}
\end{eqnarray}
where ``$\rm tr$''
indicates the trace in the space of polarization states. This
shows the relationship between the four Stokes parameters and the
$2\times 2$ density matrix $\rho$ for photon polarization states.
The density operator $\rho$ for an ensemble of free photons is given by
\begin{eqnarray}
\hat\rho=\frac{1}{\rm {tr}(\hat \rho)}\int\frac{d^3p}{(2\pi)^3}
\rho_{ij}({\bf p})a^\dagger_i({\bf p})a_j({\bf p}).
\end{eqnarray}
The parameter $I$ gives total intensity, $Q$ and $U$
intensities of linear polarizations of electromagnetic waves,
whereas the $V$ parameter measures the difference between left-
and right- circular polarizations intensities.
The expectation value of the number operator $ D^0_{ij}({\bf k})$ is defined by
\begin{eqnarray}
\langle\, D^0_{ij}({\bf k})\,\rangle\equiv {\rm tr}[\hat\rho
D^0_{ij}({\bf k})]=(2\pi)^3 \delta^3(0)(2k^0)\rho_{ij}({\bf k}).
\end{eqnarray}
The time evolution of photon polarization states is related to the
time evolution of the density matrix $\rho_{ij}(k)$, which is
governed by the following Quantum Boltzmann equation (QBE)
\cite{cosowsky1994},
\begin{eqnarray}
(2\pi)^3 \delta^3(0)(2k^0)
\frac{d}{dt}\rho_{ij}(k) = i\langle \left[H^0_I
(t);D^0_{ij}(k)\right]\rangle-\frac{1}{2}\int dt\langle
\left[H^0_I(t);\left[H^0_I
(0);D^0_{ij}(k)\right]\right]\rangle,\label{bo0}
\end{eqnarray}
where $H^0_I(t)$ is an interacting Hamiltonian. The first term on
the right-handed side is a forward scattering term, and the second
one is higher order collision term.

In our case, the interacting Hamiltonian $H^0_I(t)$ in Eq.~(\ref{bo0}) is
the Euler-Heisenberg effective Hamiltonian $H^{EH}_I(t)$
\begin{eqnarray}
H^0_I(t)=H^{EH}_I(t) = -\frac{\alpha^2}{90m_e^4}\int d^3x
\left[(F_{\mu\nu}{F}^{\mu\nu}
)^2+\frac{7}{4}(F_{\mu\nu}\tilde{F}^{\mu\nu} )^2\right],
\label{int0}
\end{eqnarray}
from Eq.~(\ref{eh}). Since $H_{EH}^I$ is the order of $\alpha^2$, in Quantum Boltzmann equation (\ref{bo0})
we approximately consider the forward scattering term only and neglect higher order collision term. The first term  $(F_{\mu\nu}{F}^{\mu\nu}
)^2\sim ({\bf E}^2-{\bf B}^2)^2$ of Eq.~(\ref{int0}) does not contribute to $\langle \left[H^0_I
(t);D^0_{ij}(k)\right]\rangle$, because its
commutation with the number operator $D^0_{ij}$ vanishes. The nontrivial  contribution comes from the term $(F_{\mu\nu}\tilde{F}^{\mu\nu} )^2\sim ({\bf E}\cdot{\bf B})^2$ in Eq.~(\ref{int0}), and $({\bf E}\cdot{\bf B})$ is odd under parity.  As a result,
the time-evolution equation for the density
matrix is approximately obtained \cite{xuei},
\begin{eqnarray}
(2\pi)^3 \delta^3(0)2k^0
\frac{d}{dt}\rho_{ij}(k) \!\!&\approx& i\langle
\left[H^{EH}_I(t),D^0_{ij}(k)\right]\rangle\nonumber\\
&=&\frac{56\alpha^2}{45m_e^4}(2\pi)^3\delta^3(0)\epsilon^{\mu\nu\mu'\nu'}\epsilon^{\alpha\beta\alpha'\beta'}k_{\alpha'}
k_\mu[\epsilon_{s\nu}(k)\epsilon_{l'\beta'}(k)]\nonumber\\
&\times&\left[\rho_{l'j}(k)\delta^{si}-\rho_{is}(k)\delta^{l'j}+\rho_{sj}(k)\delta^{l'i}-\rho_{il'}(k)\delta^{sj}\right]\nonumber\\
&\times&\!\!\! \int\frac{d^3p}{(2\pi)^32p^0}p_{\mu'}
p_{\alpha}[\epsilon_{s'\nu'}(p)\epsilon_{l\beta}(p)][\rho_{ls'}(p)+\rho_{s'l}(p)+\delta^{s'l}],
\label{j1}
\end{eqnarray}
where the following equations \cite{cosowsky1994} are used to calculate all possible contractions of creation and
annihilation operators $a_i^\dag$ and $a_j$
\begin{eqnarray}\label{contraction}
\langle \, a^\dag_{s'}(p')a_{s}(p)\, \rangle
&=&2p^0(2\pi)^3\delta^3(\mathbf{p}-\mathbf{p'})\rho_{ss'}(p),\\
\langle p|\,
a^\dag_{s'}(p')a_{s}(p)a^\dag_{l'}(q')a_{l}(q)\,|p \rangle
&=&  4p^0q^0(2\pi)^6\delta^3(\mathbf{p}-\mathbf{p'})
\delta^3(\mathbf{q}-\mathbf{q'})\rho_{ss'}(p)\rho_{ll'}(q)\nonumber\\
&+&
4p^0q^0(2\pi)^6\delta^3(\mathbf{p}-\mathbf{q'})
\delta^3(\mathbf{q}-\mathbf{p'})\rho_{s'l}(q)[\delta_{sl'}+\rho_{sl'}(p)].\label{contraction1}
\end{eqnarray}
Based on Eqs.~(\ref{v}) and (\ref{j1}),  the time-evolutions of  $V$-Stocks parameter is given by (see Ref.~\cite{xuei} for details):
\begin{eqnarray}
\dot{V}({\bf k})&=&
\hat X\Big\{Q({\bf k} )U({\bf p})\epsilon_{1\nu}({\bf k})
\epsilon_{2\beta'}({\bf k})\epsilon_{2\nu'}({\bf p})\epsilon_{1\beta}({\bf p})\nonumber\\
&+&iV({\bf k} )Q({\bf p})
\epsilon_{2\nu}({\bf k})\epsilon_{2\beta'}({\bf k})\epsilon_{1\nu'}({\bf p})\epsilon_{1\beta}({\bf p})\Big\},
\label{vd}
\end{eqnarray}
where $k$ and $p$ indicate the energy-momentum states of photons,
and the operator $\hat X$ is defined as a following integral
overall energy-momentum states $p$,
\begin{eqnarray}
\hat
X\Big\{\cdot\cdot\cdot\Big\}\equiv\frac{16\times7\alpha^2}{45m_e^4k^0}\int\frac{d^3p}{(2\pi)^32p^0}\left
[\epsilon^{\mu\nu\mu'\nu'}\epsilon^{\alpha\beta\alpha'\beta'}k_{\alpha'}
k_\mu p_{\mu'} p_{\alpha}
\right]\Big\{\cdot\cdot\cdot\Big\}.\label{int}
\end{eqnarray}
As this equation shows, the
time-evolution $\dot V$ is proportional to $Q$ and $U$ modes. This
indicates that an ensemble of linearly polarized photons will
acquire circular polarizations due to the Euler-Heisenberg
effective Lagrangian (\ref{eh}).

%%%%%%%%%%%%%%%%%%%%%%%%%%%%%%%%%%%%%%%%%%%%%%%%%%%%%%%%%%%%%%%%%%%%%%%%%%%%%%%%%%%%%%%%%%%%%%%%%%%%%%%%%%%%%%%%%%%%%%%%%%%%%%%%%%%%%%%%%
\section{Collision of two linearly polarized laser beams}
\begin{figure}
% Requires \usepackage{graphicx}
\includegraphics[width=4in]{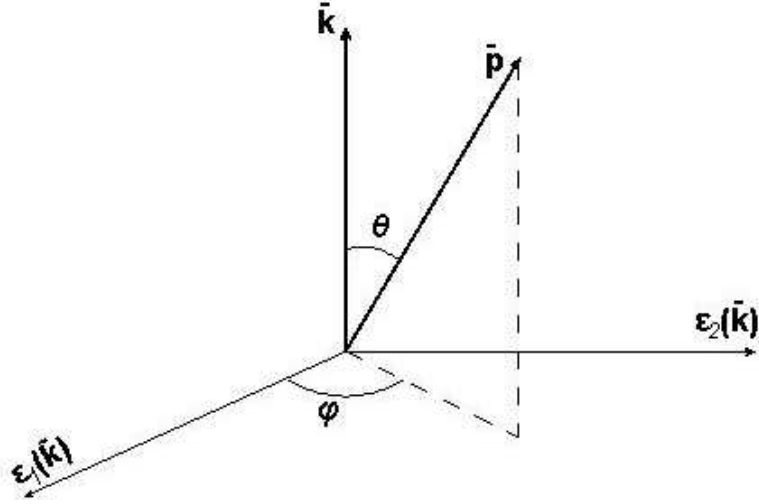}\\
\caption{This sketch shows the relative angles $\theta$ and $\phi$ between the incident laser beam direction ${\bf\bar{k}}$ and the target laser beam direction $\bar{{\bf p}}$.}\label{axis}
\end{figure}

%%%%%%%%%%%%%%%%%%%%%%%%%%%%%%%%%%%%%%%%%%%%%%%%%%%%%%%%%%%%%%%%%%%%%%%%%%%%%%%%%%%%%%%%%%%%%%%%%%%%%
Using Eq.(\ref{vd}), we calculate the circular polarization generated by the collision of two linearly polarized laser beams. In this case  the second term on the right-handed side of Eq.~(\ref{vd}) vanishes, and the density matrices of two approximately monochromatic laser beams are
\begin{equation}\label{laser1}
    \rho_{ij}({\bf p})\propto\delta^3({\bf p}-{\bf \bar{p}}),\,\,\,\,\,\,\,\rho_{ij}({\bf k})\propto\delta^3({\bf k}-{\bf \bar{k}}),
\end{equation}
where ${\bf \bar{k}}$ (${\bf \bar{p}}$) stands for the mean momentum of incident (target) laser beam, as indicated in Fig.~\ref{axis}. As a result, the integral of Eq.(\ref{int}) becomes
\begin{eqnarray}
c\int\frac{d^3p}{(2\pi)^32p^0}U({\bf p})\left
[\epsilon^{\mu\nu\mu'\nu'}\epsilon^{\alpha\beta\alpha'\beta'} p_{\mu'} p_{\alpha}
\right]=\frac{1}{2}\bar{U}(\bar{{\bf p}}) \epsilon^{\mu\nu\mu'\nu'}\,\epsilon^{\alpha\beta\alpha'\beta'}\,\hat{\bar{p}}_{\mu'} \hat{\bar{p}}_{\alpha}\label{int1}
\end{eqnarray}
where $ \hat{{\bf \bar{p}}}_{\alpha}=\bar{p}_{\alpha}/p^0$, $c$ is the speed of light and $\bar{U}(\bar{{\bf p}})$ is the mean intensity of the ``target'' laser beam. The time-evolutions (\ref{vd}) of $V$-Stocks parameter for laser beam is thus given as follows
\begin{eqnarray}
\frac{\dot{V}(\bar{{\bf k}})}{Q(\bar{{\bf k}} )}&=&\frac{8\times7\alpha^2}{45m_e^4k^0}
\bar{U}(\bar{{\bf p}})\,\epsilon^{\mu\nu\mu'\nu'}\,\epsilon^{\alpha\beta\alpha'\beta'}\,\epsilon_{1\nu}(\bar{{\bf k}})
\epsilon_{2\beta'}(\bar{{\bf k}})\epsilon_{2\nu'}(\bar{{\bf p}})\epsilon_{1\beta}(\bar{{\bf p}})\hat{\bar{p}}_{\mu'} \hat{\bar{p}}_{\alpha}\bar{k}_{\alpha'}.
\label{vd1}
\end{eqnarray}

In order to explicitly calculate Eq.~(\ref{vd1}), we coordinate $\bar{\bf k}$ in $\hat{z}$-direction,  $\vec{\epsilon}_{1}(\bar{{\bf k}})$ in $\hat{x}$-direction and $\vec{\epsilon}_{2}(\bar{{\bf k}})$ in $\hat{y}$-direction, as sketched in Fig.~\ref{axis}. In this coordinate, $\bar{\bf p}$, $\vec{\epsilon}_{1}(\bar{{\bf p}})$ and $\vec{\epsilon}_{2}(\bar{{\bf p}})$ are represented by
\begin{eqnarray}
% \nonumber to remove numbering (before each equation)
  \bar{{\bf p}} &=& \big\{\sin\theta\cos\phi,\sin\theta\sin\phi,\cos\theta\big\}\nonumber \\
  \vec{\epsilon}_{1}(\bar{{\bf p}}) &=&  \big\{\cos\theta\cos\phi,\cos\theta\sin\phi,-\sin\theta\big\}\nonumber \\
  \vec{\epsilon}_{2}(\bar{{\bf p}}) &=&  \big\{-\sin\phi,\cos\phi,0\big\}.\label{condition1}
\end{eqnarray}
in terms of polar angles $\theta$ and $\phi$ (see Fig.\ref{axis}), and indexes
\begin{equation}\label{condition2}
    \nu=1;\,\,\,\,\,\beta'=2;\,\,\,\,\,\,\mu=0,3;\,\,\,\,\,\,\alpha'=0,3,
\end{equation}
in Eq.~(\ref{vd1}).
As a result we calculate Eq.~(\ref{vd1}), yielding
\begin{eqnarray}
\frac{\dot{V}(\bar{{\bf k}})}{Q(\bar{{\bf k}} )}&=&\frac{56\alpha^2}{45m^4k^0}
\bar{U}(\bar{{\bf p}})\,\epsilon^{\mu\nu\mu'\nu'}\,\epsilon^{\alpha\beta\alpha'\beta'}\,\epsilon_{1\nu}(\bar{{\bf k}})
\epsilon_{2\beta'}(\bar{{\bf k}})\epsilon_{2\nu'}(\bar{{\bf p}})\epsilon_{1\beta}(\bar{{\bf p}})\hat{\bar{p}}_{\mu'} \hat{\bar{p}}_{\alpha}\bar{k}_{\alpha'}
\bar{k}_\mu\nonumber\\
&\times&\delta^{1\nu}\delta^{2\beta'}\big\{\delta^{0\mu}\delta^{0\alpha'}\big[\delta^{3\alpha}(\delta^{2\mu'}+
\delta^{3\mu'})+\delta^{1\alpha}(\delta^{2\mu'}+\delta^{3\mu'})\big]+\delta^{3\mu}\delta^{3\alpha'}\delta^{0\mu'}\delta^{0\alpha}\nonumber\\
&+&\delta^{0\mu}\delta^{3\alpha'}\delta^{0\alpha}(\delta^{2\mu'}+
\delta^{3\mu'})+\delta^{3\mu}\delta^{0\alpha'}\delta^{0\mu'}(\delta^{1\alpha}+\delta^{3\alpha})\big\}\nonumber\\
&=&\frac{56\alpha^2}{45m^4k^0}
\bar{U}(\bar{{\bf p}})\,(k^0)^2\Big\{\vec{\epsilon}_{1k}\cdot\vec{\epsilon}_{1p}\,\hat{{\bf\bar{p}}}\cdot\vec{\epsilon}_{2k}\,
\hat{{\bf\bar{k}}}\cdot\vec{\epsilon}_{2p}\left[\hat{{\bf\bar{p}}}\cdot\hat{{\bf\bar{k}}}-1\right]
\nonumber\\
&-&\vec{\epsilon}_{1k}\cdot\vec{\epsilon}_{1p}\,\vec{\epsilon}_{2k}\cdot\vec{\epsilon}_{2p}\,
\left[1+(\hat{{\bf\bar{p}}}\cdot\hat{{\bf\bar{k}}})^2-2\hat{{\bf\bar{p}}}\cdot\hat{{\bf\bar{k}}}\right]-
\vec{\epsilon}_{2k}\cdot\vec{\epsilon}_{2p}\,\hat{{\bf\bar{p}}}\cdot\vec{\epsilon}_{1k}\,
\hat{{\bf\bar{k}}}\cdot\vec{\epsilon}_{1p}\nonumber\\
&+&\hat{{\bf\bar{p}}}\cdot\hat{{\bf\bar{k}}}\,\vec{\epsilon}_{2k}\cdot\vec{\epsilon}_{2p}
\,\hat{{\bf\bar{p}}}\cdot\vec{\epsilon}_{1k}\,\hat{{\bf\bar{k}}}\cdot\vec{\epsilon}_{1p}-\hat{{\bf\bar{p}}}\cdot\vec{\epsilon}_{2k}
\,\hat{{\bf\bar{k}}}\cdot\vec{\epsilon}_{2p}\,\hat{{\bf\bar{p}}}\cdot\vec{\epsilon}_{1k}\,\hat{{\bf\bar{k}}}\cdot\vec{\epsilon}_{1p}\Big\},
\label{vd3}
\end{eqnarray}
where $\vec{\epsilon}_{ip}\equiv \vec{\epsilon}_{i}(\bar{{\bf p}})$, $\vec{\epsilon}_{ik}\equiv \vec{\epsilon}_{i}(\bar{{\bf k}})$ and
$i=1,2$. Substituting Eq.~(\ref{condition1}) into Eq.~(\ref{vd3}), we obtain the final result
\begin{eqnarray}
\frac{\dot{V}(\bar{{\bf k}})}{Q(\bar{{\bf k}} )}&=&\frac{56}{45}(\frac{k^0}{m})^2
\frac{\bar{U}(\bar{{\bf p}})}{k^0}\,\alpha^2 \lambda_e^2\,\cos^2\phi\left[1+\cos^2\theta-2\cos\theta\right],
\label{vd4}
\end{eqnarray}
which is maximal for the head-head collision of two laser beams ($\theta=\pi$ and $\phi=0$), at given intensities $Q(\bar{{\bf k}} )$ and $\bar{U}(\bar{{\bf p}})$ of two linearly polarized laser beams.

To end this section, we recall
the QED birefringence \cite{adler1971,hh1997} of two photon polarizations being different in their propagating directions, intensities and phases in an external magnetic field, possibly leading to circular or elliptical polarizations.
In the collision of two linearly polarized laser beams, we show that due to the term $({\bf E}\cdot{\bf B})^2$ in Eq.~(\ref{int0}), the time-averaged Stokes parameter $V$ for circular polarizations does not vanishes, and evolves with the interacting time of two laser beams.

%%%%%%%%%%%%%%%%%%%%%%%%%%%%%%%%%%%%%%%%%%%%%%%%%%%%%%%%%%%%%%%%%%%%%%%%%%%%%%%%%%%%%%%%%%%%%%%%%%%%%%%%%%%%%%%%%%%%%%%%%%%%%%%%%%%%%%%%%
\section{The rate of generating laser photons with circular polarization}
%%%%%%%%%%%%%%%%%%%%%%%%%%%%%%%%%%%%%%%%%%%%%%%%%%
We estimate the rate of circular polarization generation by the collision of  two linearly polarized laser beams with a mean energy $\bar k^0=\bar p^0$ and average power $\bar{P}$,
\begin{equation}\label{beam1}
    \bar{k}^0=\bar{p}^0\sim {\rm eV},\,\,\,\,\,\,\,\,\bar{P}^{in}({\bf\bar{p}})=\bar{P}^{T}({\bf\bar{k}})=\bar{P}\sim {\rm KW}.
\end{equation}
Suppose that $A$ is the cross-sectional interacting area of two laser beams, $\Delta t$ is the interacting time of two laser beams, and within the interacting time $\Delta t$, two laser beams have continuous beam profile.
In this case Eq.(\ref{vd4}) can be written as
\begin{eqnarray}
\frac{\Delta V(\bar{{\bf k}})}{\bar{k}^0}&\simeq&1.25\,\Delta t\,\,\alpha^2 \lambda_e^2\,\cos^2\phi\left[1+\cos^2\theta-2\cos\theta\right](\frac{\bar{k}^0}{m})^2
\left(\frac{\bar{P}^{in}}{A\,\bar{k}^0}\right)\left(\frac{\bar{P}^{T}}{A\,\bar{k}^0}\right).
\label{vd52}
\end{eqnarray}
Because  $\Delta V(\bar{{\bf k}})$ is the intensity of circularly polarized laser photons with the energy $\bar k^0$, the average rate $R_V$ of generating circularly polarized laser photons (number /sec) after two beams interacting for the time-interval $\Delta t$ is given by
\begin{eqnarray}
R_V&\simeq&1.25\,\Delta t\,\,\alpha^2 \lambda_e^2\,\cos^2\phi\left[1+\cos^2\theta-2\cos\theta\right](\frac{\bar{k}^0}{m})^2
\left(\frac{\bar{P}^{in}}{A\,\bar{k}^0}\right)\left(\frac{\bar{P}^{T}}{\,\bar{k}^0}\right)\,,
\label{rate01}
\end{eqnarray}
where the electron Compton length $\lambda_e\simeq 3.86\times 10^{-11}{\rm cm}$.
As an example, using two laser beams, $\bar{k}^0\sim {\rm eV}$, $\Delta t\sim 0.3 \,\mu\, {\rm sec}$, $\bar{P}\sim {\rm KW}$ and $A\sim 1\,{\rm cm^2}$, which are available in laboratories, we obtain the rate $R_V\simeq 200/{\rm sec}$ that should be large enough to be measurable.
Note that the rate $R_V$ of Eq.~(\ref{rate01}) is proportional to
$\alpha^2$, compared with the rate $R_{\gamma\gamma}$ of Eq.~(\ref{rate0}), which is proportional to $\alpha^4$. The reason is that the leading contribution to the generation of circular polarization comes from the forward scattering term of Eq.~(\ref{bo0}), which is the order of $\alpha^2$. 
This explains why the measurable rate $R_V$ of
Eq.~(\ref{rate01}) needs much less power (KW) of laser beams, than petawatt (PW) of laser beams for a sizable rate $R_{\gamma\gamma}$.

\comment{The above equation indicates to observe a photon with circular polarization per second due to photon-photon interactions, at least two linearly polarized laser beam with average power about $0.1{\rm KW}$, beam diameter about $\sqrt{A}\sim1{\rm cm}$  and average wavelength about $\sim1{\rm \mu m}$ is needed.}

%%%%%%%%%%%%%%%%%%%%%%%%%%%%%%%%%%%%%%%%%%%%%%%%%%%%%%%%%%%%%%%%%%%%%%%%%%%%%%%%%%%%%%%%%%%%%%%%%%%%%%%%%%%%

\section{Conclusion and Remarks}
To study photon-photon interactions of quantum electrodynamics, we approximately calculate the circular polarization intensity  generated by the collision between two linearly polarized laser beams, and obtain the result of Eq.~(\ref{vd52}).
For this purpose, we approximately solve the Quantum
Boltzmann Equation for the density matrix of photon ensemble with the nonlinear Euler-Heisenberg effective Lagrangian, and obtain the
time-evolution of Stokes parameter $V$ for circular polarization.
%We explain the effect that the colliding of two laser beams converts their linear polarizations to circular polarizations.
Using some parameters of available KW laser beams in laboratories,
we estimate the rate (\ref{rate01}) of generating circularly polarized
photons (number/sec), which seems to be large enough for possible measurements. How to experimentally measure the circular polarization generated is not in the scope of this Letter.

The phase shift due to the nonlinear interactions of ultra-intense (PW) laser beams and some sensitive techniques to detect this phase shift have been studied in Refs.~\cite{rev4,rev7}. The power of laser beams which needs to measure this phase-shift is about four order of magnitude larger than the power of laser beams used to produce measurable circular polarizations.
The proposed ELI project \cite{eli} will provide laser pulses of wavelength $\sim 800$nm, intensity $ 10^{29}{\rm WM}^{-2}$ (with peak power $\sim$PW and average power $\sim3$MW), the repetition rate of pulses $f_{\rm pulse}\sim 10{\rm Hz}$ and time-duration of a pulse $\sim 30\,$fs, and the size of focusing spot $\sim 10 {\rm \mu m}$. This laser facility will make it be possible to detect the phase shift, circular polarization and other effects originated from the nonlinear photon-photon interactions in quantum electrodynamics.

%\section*{\small Acknowledgment}

%%%%%%%%%%%%%%%%%%%%%%%%%%%%%%%%%%%%%%%%%%%%%%%%%%%%%%%%%%%%%%%%%%%%%%%%%%%%%%%%%%

\end{document}